\documentclass[aps,pre]{revtex4}
\usepackage{wrapfig}
\usepackage{url}
\usepackage{amssymb,amstext,amsmath}
\usepackage{dsfont}
\usepackage[T1]{fontenc}
\usepackage{graphicx}
\usepackage{stmaryrd}
\usepackage{hyperref}

\newcommand{\eq}[1]{\begin{align}#1\end{align}}

\DeclareMathAlphabet{\varmathbb}{U}{bbold}{m}{n}

\let\Horig\H
\renewcommand{\H}{{\rm H}}

\newcommand{\E}{\mathbb{E}}

\begin{document}

\title{Matrix Completion from Fewer Entries:\\
Spectral Detectability and Rank Estimation}

\author{A. Saade$^1$, F. Krzakala$^{1,2,3}$ and L. Zdeborov\'a$^4$}

\affiliation{$^1$ Laboratoire de Physique Statistique, UMR 8550 CNRS,  Department of Physics, {\'E}cole Normale Sup{\'e}rieure and PSL  Research University, Rue Lhomond, 75005  Paris, France \\
  $^2$ Sorbonne Universit\'es, UPMC Univ Paris 06, UMR 8550, LPS,  F-75005, Paris, France\\ $^3$ ESPCI and CNRS UMR 7083
Gulliver, 10 rue Vauquelin,Paris 75005\\ $^4$ Institut de Physique
Th\'eorique, CEA Saclay and URA 2306, CNRS, 91191 Gif-sur-Yvette,
France}


\begin{abstract}
The completion of low rank matrices from few entries is a task with many practical applications. We consider here two aspects of this problem: detectability, i.e. the ability to estimate the rank $r$ reliably from the fewest possible random entries, and performance in achieving small reconstruction error. We propose a spectral algorithm for these two tasks called MaCBetH (for Matrix Completion with the Bethe Hessian). The rank is estimated as the number of negative eigenvalues of the Bethe Hessian matrix, and the corresponding eigenvectors are used as initial condition for the minimization of the discrepancy between the estimated matrix and the revealed entries.  We analyze the performance in a random matrix setting using results from the statistical mechanics of the Hopfield neural network, and show in particular that MaCBetH efficiently detects the rank~$r$ of a large $n\times m$ matrix from $C(r)r\sqrt{nm}$ entries, where $C(r)$ is a constant depicted in Fig.~\ref{fig:transition}.  We also evaluate the corresponding root-mean-square error empirically and show that MaCBetH compares favorably to other existing approaches.  \end{abstract}

\date{\today}
\maketitle

\section{Introduction}
Matrix completion is the task of inferring the missing entries of a matrix given a subset of known entries. Typically, this is possible because the matrix to be completed has (at least approximately) low rank $r$. This problem has witnessed a burst of activity, see e.g. \cite{candes2009exact,candes2010power,kmo10}, motivated by many applications such as collaborative filtering \cite{candes2009exact}, quantum tomography \cite{gross2010quantum} in physics, or the analysis of a covariance matrix \cite{candes2009exact}.A commonly studied model for matrix completion assumes the matrix to be exactly low rank, with the known entries chosen uniformly at random and observed without noise. The most widely considered question in this setting is how many entries need to be revealed such that the matrix can be completed exactly in a computationally efficient way \cite{candes2009exact,kmo10}. While our present paper assumes the same model, the main questions we investigate are different.

\medskip
The first question we address in this paper is {\it detectability}, i.e. how many random entries do we need to reveal in order to be able to estimate the rank $r$ reliably. This question is motivated by the more generic problem of detecting structure (in our case, low rank) hidden in partially observed data.  It is reasonable to expect the existence of a region where exact completion is hard or even impossible yet the rank estimation is tractable. A second question we address is what is the minimum achievable root-mean-square error (RMSE) in estimating the unknown elements of the matrix. In practice, even if exact reconstruction is not possible, having a procedure that provides a very small RMSE might be quite sufficient. 

\medskip
In this paper we propose an algorithm called MaCBetH that gives the best known empirical performance for the two tasks above when the rank $r$ is small. The rank in our algorithm is estimated as the number of negative eigenvalues of an associated Bethe Hessian matrix \cite{saade2014spectral,saade2015spectral}, and the corresponding eigenvectors are used as an initial condition for the local optimization of a cost function commonly considered in matrix completion (see e.g. \cite{kmo10}). In particular, in the random matrix setting, we show that MaCBetH detects the rank of a large $n\times m$ matrix from $C(r) r \sqrt{nm}$ entries, where $C(r)$ is a small constant, see Fig.~\ref{fig:transition}, and $C(r) \to 1$ as $r\to \infty$. The corresponding RMSE is evaluated empirically, and in the regime close to $C(r) r \sqrt{nm}$, it compares very favorably to existing approaches, in particular to OptSpace \cite{kmo10}.

\medskip
This contribution is organized as follows. First, in Sec. \ref{sec:def} we define the problem and present generally our approach in the context of existing works. In Sec. \ref{sec:algo} we describe our algorithm and motivate its construction via a spectral relaxation of the Hopfield model of neural network. 
Next, in Sec. \ref{sec:analysis} we show how the performance of the proposed spectral method can be analyzed using, in parts, results from spin glass theory and phase transitions, and rigorous results on the spectral density of large random matrices. Finally, in Sec.~\ref{sec:numerical} we present numerical simulations that demonstrate the efficiency of MaCBetH.

\section{Problem definition and relation to other works}
\label{sec:def}

Let ${\cal M}^{\rm true}$ be a rank-$r$ matrix such that
\eq{
{\cal M}^{\rm true}= X  Y^{\dag} \, ,
}
where $X\in\mathbb{R}^{n\times r}$ and $Y\in\mathbb{R}^{m\times r}$ are two (unknown) tall matrices. We observe only a small fraction of 
the elements of ${\cal M}^{\rm true}$, chosen uniformly at random. We call $E$
the subset of observed entries, and ${\cal M}$ the (sparse) matrix supported on $E$ whose nonzero elements are the revealed entries of ${\cal M} ^{\rm true}$.  
The aim is to reconstruct the rank $r$ matrix ${\cal M} ^{\rm true}=XY^{\dag}$ given 
${\cal M}$. An important parameter which controls the difficulty of the problem is $\epsilon = |E|/\sqrt{nm}$. In the case of a square matrix ${\cal M}$, this is the average number of revealed entries per line or column.

In our numerical examples and theoretical justifications we shall generate the low rank matrix $ {\cal M}^{\rm true}= X Y^{\dag} $, using tall matrices $X$ and $Y$ with iid Gaussian elements, we call this the random matrix setting. The MaCBetH algorithm is, however, non-parametric and does not use any prior knowledge about $X$ and $Y$. The analysis we perform applies to the limit $n\to \infty$ while $m/n=\alpha=O(1)$ and $r=O(1)$. 

The matrix completion problem was popularized in \cite{candes2009exact} who proposed nuclear norm minimization as a convex relaxation of the problem. The algorithmic complexity of the associated semidefinite programming is, however, $O(n^2m^2)$.
A low complexity procedure to solve the problem was later proposed by \cite{cai2010singular} and is based on singular value decomposition (SVD). A considerable step towards theoretical understanding of matrix completion from few entries was made in \cite{kmo10} who proved that with the use of {\it trimming} the performance of SVD-based matrix completion can be improved and a RMSE proportional to $\sqrt{nr/|E|}$ can be achieved. The algorithm of \cite{kmo10} is referred to as OptSpace, and empirically it achieves state-of-the-art RMSE in the regime of very few revealed entries.  

OptSpace proceeds in three steps \cite{kmo10}. First, one trims the observed matrix ${\cal M}$ by setting to zero all rows (resp. columns) with more revealed entries than twice the average number of revealed entries per row (resp. per column). 
Second, a singular value decompositions is performed on the matrix and only the first $r$ components are kept. When the rank $r$ is unknown it is estimated as the index for which the ratio between two consecutive singular values has a minimum.  Third, a local minimization of the discrepancy between the observed entries and a low-rank estimate is performed. The initial condition for this minimization is given by the first $r$ left and right singular vectors from the second step. 

In this work we improve upon OptSpace by replacing the first two steps by a different spectral procedure that detects the rank and provides a better initial condition for the discrepancy minimization. Our method leverages on recent progress made in the task of detecting communities in the stochastic block model \cite{krzakala2013spectral,saade2014spectral} with spectral methods. Both in community detection and matrix completion, traditional spectral methods fail in the very sparse regime due to the existence of spurious large eigenvalues (or singular values) corresponding to localized eigenvectors \cite{krzakala2013spectral,kmo10}. The authors of \cite{krzakala2013spectral,saade2014spectral,blm15} showed that using the non-backtracking matrix or the closely related Bethe Hessian as a basis for the spectral method in community detection provides reliable rank estimation and better inference performance. The present paper provides an analogous improvement for the matrix completion problem. In particular, we shall analyze the algorithm using tools from spin glass theory in statistical mechanics, and show that there exists a phase transition between a phase where it is able to detect the rank, and a phase where it is unable to do so.

\section{Algorithm and motivation}
\label{sec:algo}

\subsection{The MacBetH algorithm}
\begin{figure}[t] \begin{center} \includegraphics[width=1\linewidth]{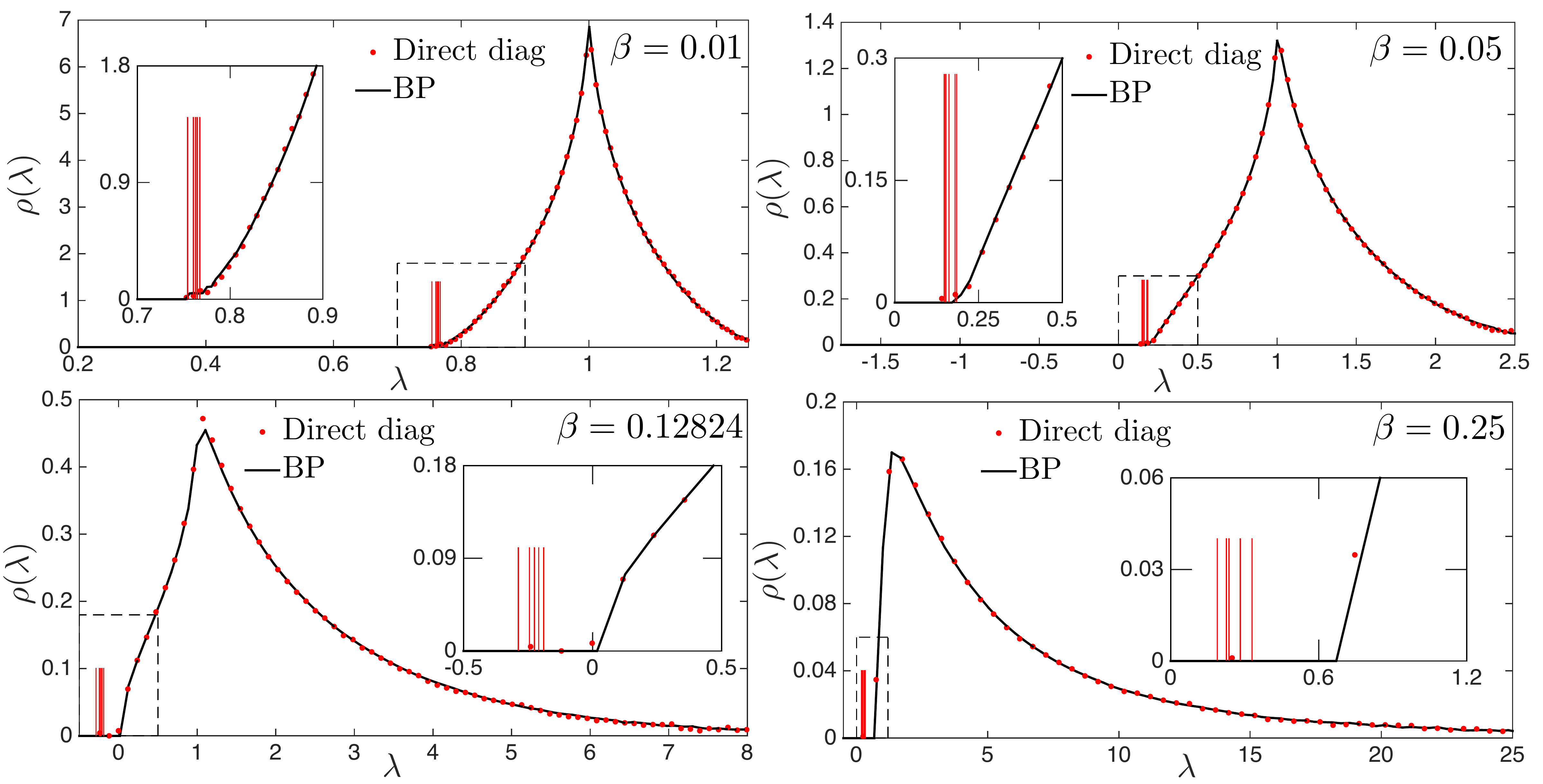} \end{center} \caption{ Spectral density of the Bethe Hessian for various values of the parameter $\beta$.  The red dots are the result of the direct diagonalisation of the Bethe Hessian for a rank $r=5$ and $n=m=10^4$ matrix, with $\epsilon = 15$ revealed entries per row on average. The black curves are the solutions to the recursion (\ref{BPrec}) computed with belief propagation on a graph of size $10^5$.  We isolated the 5 smallest eigenvalues, represented as small bars for convenience, and the inset is a zoom around these smallest eigenvalues. For $\beta$ small enough (top plots), the Bethe Hessian is positive definite, signaling that the paramagnetic state (\ref{paramagnetic_state}) is a local minimum of the Bethe free energy. As $\beta$ increases, the spectrum is shifted towards the negative region and has 5 negative eigenvalues at the approximate value of $\hat \beta_{\rm SG}=0.12824$ (to be compared to $\beta_R=0.0832$
for this case) evaluated by our algorithm (lower left plot). These eigenvalues, corresponding to the retrieval states (\ref{retrieval_states}), become positive and eventually merge in the bulk as $\beta$ is further increased (lower right plot), while the bulk of uninformative eigenvalues remains at all values of $\beta$ in the positive region.  \label{fig:spectrum}} \end{figure}

A standard approach to the completion problem (see e.g.~\cite{kmo10}) is to minimize the cost function 
\eq{ \underset{X,Y}{\min}  \sum_{(ij)\in E} [{\cal M}_{ij} - (X Y^{\dag} )_{ij} ]^2 
\label{costFunction}
}
over $X\in\mathbb{R}^{n\times r}$ and $Y\in\mathbb{R}^{m\times r}$. This function is non-convex, and global optimization is hard. One therefore resorts to a local optimization technique with a careful choice of 
the initial conditions $X_0,Y_0$.
In our method, given the matrix ${\cal M}$, we consider a weighted bipartite undirected graph with adjacency matrix $A\in\mathbb{R}^{(n+m)\times (n+m)}$
\eq{
A=\left( \begin{array}{cc}
0 & {\cal M} \\
{\cal M}^{T} & 0
\end{array} \right)\,.
}
We will refer to the graph thus defined as $\mathcal{G}$. We now define the Bethe Hessian matrix $\H(\beta)\in\mathbb{R}^{(n+m)\times (n+m)}$ to be the matrix with elements 
\eq{
\H_{ij}(\beta) = \left( 1 + \underset{ k\in\partial i }{ \sum }  \sinh^2{ \beta A_{ik}}  \right) \delta_{ij} - \frac{1}{2}\sinh(2\beta A_{ij})\, ,
\label{BetheHessian_def}
}
where $\beta$ is a parameter that we will fix to a well-defined value $\beta_{\rm SG}$ depending on the data, and $\partial i $ stands for the neighbors of $i$ in the graph $\mathcal{G}$. The MaCBetH algorithm that is the main subject of this paper is then, given the matrix $A$, which we assume to be centered:

\bigskip 
\textbf{Algorithm (MaCBetH)}
\begin{enumerate}
\item Numerically solve for the value of $\hat{\beta}_{\rm SG}$ such that 
\eq{\displaystyle F(\hat{\beta}_{\rm SG}) \equiv \frac{1}{\sqrt{nm}}\underset{(i,j)\in E}{\sum} \tanh^2(\hat{\beta}_{\rm SG}{\cal M}_{ij}) = 1\, . \label{eq:hat_sg}}  
%
\item Build the Bethe Hessian $\H(\hat{\beta}_{\rm SG})$ following eq. (\ref{BetheHessian_def}).  
\item Compute {\it all} its negative eigenvalues $\lambda_1,\cdots,\lambda_{\hat{r}}$ and corresponding eigenvectors $v_1,\cdots,v_{\hat{r}}$. ${\hat{r}}$ is our estimate for the rank $r$. Set $X_0$ (resp. $Y_0$) to be the first $n$ lines (resp. the last $m$ lines) of the matrix $[v_1\ v_2\ \cdots\ v_{\hat{r}}]$.
\item Perform local optimization of the cost function (\ref{costFunction}) with rank $\hat{r}$ and initial condition $X_0,Y_0$.
\end{enumerate}
The function $F$, in the first step, being an increasing function of $\beta$, $\hat{\beta}_{\rm SG}$ can be found efficiently, e.g. by dichotomy. Alternatively, $\hat \beta_{SG}$ in step 1 can be tuned in such a way that the number of negative eigenvalues of the Bethe Hessian is the largest possible. In step 2 we could also use the non-backtracking matrix weighted by $\tanh{\beta{\cal M}_{ij}}$, it was shown in \cite{saade2014spectral} that the spectrum of the Bethe Hessian and the non-backtracking matrix are closely related. 
In the next section, we will motivate and analyze this algorithm (in the setting where ${\cal M}^{\rm true}$ was generated from elements-wise random $X$ and $Y$) and show that in this case  MaCBetH is able to infer the rank whenever $\epsilon>\epsilon_c$. Fig.~\ref{fig:spectrum} illustrates the spectral properties of the Bethe Hessian that justify this algorithm: the spectrum is composed of a few informative negative eigenvalues, well separated from the bulk (which remains positive). This algorithm is computationally efficient as it is based on the eigenvalue decomposition of a sparse, symmetric matrix.

\subsection{Motivation from a Hopfield model}
We shall now motivate the construction of the MaCBetH algorithm from a graphical model perspective and a spectral relaxation. Given the observed matrix $\cal M$ from the previous section, we consider the following graphical model
\eq{
P(\{s\},\{t\}) = \frac{1}{Z}\exp{\left( \beta \underset{(i,j)\in E}\sum
{\cal M}_{ij}s_it_j\right) }\, ,
\label{jpd}
}
where the $\{s_i\}_{1\leq i\leq n}$ and $\{t_j\}_{1\leq j\leq m}$ are binary variables, and $\beta$ is a parameter controlling the strength of the interactions. This model is a (generalized) Hebbian Hopfield model \cite{hopfield1982neural} on a bipartite sparse graph.  To study it, we can use the standard Bethe approximation which is widely believed to be exact for such problems on large random graphs \cite{yedidia2001bethe,MM2009}. In this approximation the means $\E(s_i),\E(t_j)$ and moments $\E(s_it_j)$ of each variable are approximated by
the parameters $b_i,c_j$ and $\xi_{ij}$ that minimize the so-called Bethe
free energy $F_{\rm Bethe}(\{b_i\},\{c_j\},\{\xi_{ij}\})$ that reads
\eq{
\label{freeEnergy}
\notag &F_{\rm Bethe}(\{b_i\},\{c_j\},\{\xi_{ij}\})=-\underset{(i,j)\in E}{\sum}{\cal M}_{ij}\xi_{ij}+\underset{(i,j)\in E}{\sum}\ \underset{s_i,t_j}{\sum}\eta\Big(\frac{1+b_is_i+c_jt_j+\xi_{ij}s_it_j}{4}\Big)\\
&+\overset{n}{\underset{i=1}{\sum}}(1-d_i)\underset{s_i}{\sum}\eta\Big(\frac{1+b_is_i}{2}\Big)+\overset{m}{\underset{j=1}{\sum}}(1-d_j)\underset{t_j}{\sum}\eta\Big(\frac{1+c_jt_j}{2}\Big)\, , } where $\eta(x):=x\ln{x}$, and $d_i,d_j$ are the degrees of nodes $i$ and $j$ in the graph $\mathcal{G}$. 

Neural networks models such as eq.~(\ref{jpd}) have been extensively studied over the last decades (see e.g. \cite{amit1985spin,wemmenhove2003finite,castillo2004little,MM2009,zhang2015nonbacktracking} and references therein) and the phenomenology, that we shall review briefly here, is well known. In particular, for $\beta$ small enough, the global minimum of the Bethe free energy corresponds to the so-called \emph{paramagnetic state} 
\eq{
\forall i,j,\qquad b_i = c_j = 0,\ \xi_{ij} = \tanh{(\beta {\cal M}_{ij})}.
\label{paramagnetic_state}
}
As we increase $\beta$, above a certain value $\beta_{\rm R}$, the model enters a \emph{retrieval} phase, where the free energy has local minima correlated with the factors $X$ and $Y$. There are $r$ local minima, called \emph{retrieval states} $(\{b_i^l\},\{c_j^l\},\{\xi_{ij}^l\})$ indexed by $l =1,\cdots, r$ such that, in the large $n,m$ limit,
\eq{
\qquad \forall l=1\cdots r,\qquad \frac{1}{n}\overset{n}{\underset{i=1}{\sum}} X_{i,l}b_i^l > 0,\quad \frac{1}{m}\overset{m}{\underset{j=1}{\sum}} Y_{j,l}c_j^l > 0\, .
\label{retrieval_states}
}
These retrieval states are therefore well-suited as initial conditions for the local optimization of eq.~(\ref{costFunction}), and we expect their number to tell us the correct rank. Increasing $\beta$ above a critical value $\beta_{\rm SG}$ the system eventually enters a spin glass phase, marked by the appearance of many spurious minima.

It would be tempting to continue the Bethe approach and to derive the belief propagation equations, but we shall here instead consider a simpler spectral relaxation of the problem, following the same strategy as used in \cite{saade2014spectral,saade2015spectral} for graph clustering. First, we use the fact that the paramagnetic state (\ref{paramagnetic_state}) is always a stationary point of the Bethe free energy, for any value of $\beta$ \cite{mooij2004validity,ricci2012bethe}. In order to detect the retrieval states, we thus study its stability by looking for negative eigenvalues of the Hessian of the Bethe free energy evaluated at the paramagnetic state (\ref{paramagnetic_state}). At this point, the elements of the Hessian involving one derivative with respect to $\xi_{ij}$ vanish, while the block involving two such derivatives is a diagonal positive definite matrix \cite{saade2014spectral,mooij2004validity}. The remaining part is the matrix called Bethe Hessian in \cite{saade2014spectral}, and eigenvectors corresponding to its negative eigenvalues are thus expected to give an approximation of the retrieval states (\ref{retrieval_states}). The picture exposed in this section is summarized in figure \ref{fig:spectrum}. This motivates the use of the MaCBetH algorithm.

Note that a similar approach was used in \cite{zhang2015nonbacktracking} to detect the retrieval states of a Hopfield model using the weighted non-backtracking matrix \cite{krzakala2013spectral}, which linearizes the belief propagation equations rather than the Bethe free energy, resulting in a larger, non-symmetric matrix. The Bethe Hessian, while mathematically closely related, is also simpler to handle in practice.

\section{Analysis of performance in detection}
\label{sec:analysis}
We now show how the performance of MaCBetH can be analyzed, and the spectral properties of the matrix characterized using both tools from statistical mechanics and rigorous arguments.

\subsection{Analysis of the phase transition}
\label{analysis}
We start by investigating the phase transition above which our spectral method will detect the correct rank. Let $x_p=(x_p^{l})_{1\leq l\leq r},y_p=(y_p^{l})_{1\leq l\leq r}$ be random vectors with the same empirical distribution as the lines of $X$ and $Y$ respectively. Using the statistical mechanics correspondence between the negative eigenvalues of the Bethe Hessian and the appearance of phase transitions in model (\ref{jpd}), we can compute the values $\beta_{R}$ and $\beta_{\rm SG}$ where instabilities towards, respectively, the retrieval states and the spurious glassy states, arise. We have repeated the computations of \cite{amit1985spin,wemmenhove2003finite,castillo2004little,zhang2015nonbacktracking} in the case of model (\ref{jpd}), using the cavity method \cite{MM2009}. 
We refer the reader interested in the technical details of the statistical mechanics 
approach to neural networks to \cite{wemmenhove2003finite,castillo2004little,zhang2015nonbacktracking}.

Following a standard computation for locating phase transitions in the Bethe approximation (see e.g. \cite{MM2009,zdeborova2009statistical}), the stability of the paramagnetic state (\ref{paramagnetic_state}) towards these two phases can be monitored in terms of the two following parameters:
\eq{
\label{lambda}
\lambda(\beta) &= \underset{s\rightarrow\infty}{\lim}\ \E\Big[
\overset{s}{\underset{p=1}{\prod}}\tanh^2\Big(\beta \overset{r}{\underset{l=1}{\sum}}x_p^{l}y_p^{l}\Big) \tanh^2\Big(\beta \overset{r}{\underset{l=1}{\sum}}x_{p+1}^{l}y_p^{l}\Big)
\Big]^\frac{1}{2s}\, ,\\	
\mu(\beta) &= \underset{s\rightarrow\infty}{\lim} \ \E\Big[
\overset{s}{\underset{p=1}{\prod}}\tanh\Big(\beta|x_p^{1}y_p^{1}|+\beta \overset{r}{\underset{l=2}{\sum}}x_p^{l}y_p^{l}\Big) \tanh\Big(\beta|x_{p+1}^{1}y_p^{1}|+\beta\overset{r}{\underset{l=2}{\sum}}x_{p+1}^{l}y_p^{l}\Big)
\Big]^\frac{1}{2s}\, ,
\label{mu}
} where the expectation is over the distribution of the vectors $x_p,y_p$. The parameter $\lambda(\beta)$ controls the sensitivity of the paramagnetic solution to random noise, while $\mu(\beta)$ measures its sensitivity to a perturbation in the direction of a retrieval state. $\beta_{\rm SG}$ and $\beta_{R}$ are defined implicitly as $\epsilon \lambda(\beta_{\rm SG})=1$ and $\epsilon \mu(\beta_{R}) =1$, i.e. the value beyond which the perturbation diverges. The existence of a retrieval phase is equivalent to the condition $\beta_{\rm SG} > \beta_R$, so that there exists a range of values of $\beta$ where the retrieval states exist, but not the spurious ones. If this condition is met, by setting $\beta = \beta_{SG}$ in our algorithm, we ensure the presence of meaningful negative eigenvalues of the Bethe Hessian.

\begin{figure}
\centering
\includegraphics[width=0.8\linewidth]{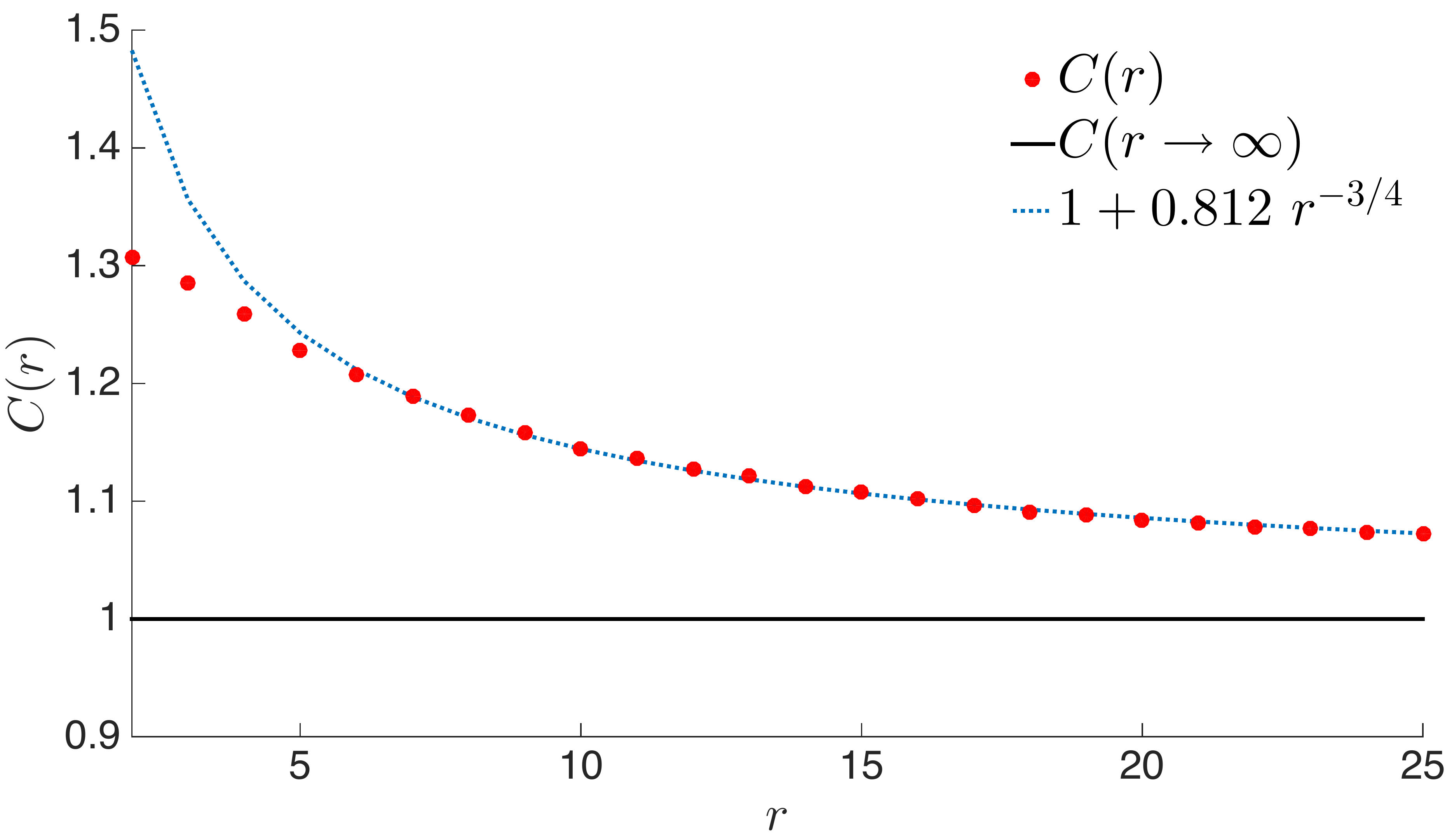}
\caption{Location of the critical value as a function of the rank $r$. MaCBetH is able to estimate the correct rank from $|E| > C(r) r \sqrt{nm}$ known entries. We used a population dynamics algorithm with a population of size $1\times10^6$ to compute the functions $\lambda$ and $\mu$ from (\ref{lambda},\ref{mu}). The dotted line is a fit suggesting that $C(r)-1 = O(r^{-3/4})$. 
\label{fig:transition}}
\end{figure}

We define the critical value of $\epsilon=\epsilon_{c}$ such that $\beta_{\rm SG} > \beta_R$ if and only if $\epsilon>\epsilon_{c}$. In general, there is no closed-form formula for this critical value, which is defined implicitly in terms of the functions $\lambda$ and $\mu$. We thus computed $\epsilon_{c}$ numerically using a population dynamics algorithm~\cite{MM2009} and the results for $C(r) = \epsilon_c/r$ are presented on Figure \ref{fig:transition}. Quite remarkably, with the definition $\epsilon = |E|/\sqrt{n m}$, the critical value $\epsilon_c$ does not depend on the ratio $m/n$, only on the rank $r$.

In the limit of large $\epsilon$ and $r$ it is possible to obtain a simple closed-form formula. In this case the observed entries of the matrix become jointly Gaussian distributed, and uncorrelated, and therefore independent. Expression (\ref{lambda}) then simplifies to 
\eq{ 
\lambda(\beta) \underset{r\to\infty}{\sim} \E\Big[\tanh^2\Big(\beta \overset{r}{\underset{l=1}{\sum}}x^{l}y^{l}\Big) \Big]\, .  } 
We stress that the MaCBetH algorithm uses an empirical estimator (\ref{eq:hat_sg}) of this quantity to compute an approximation $\hat \beta_{\rm SG}$ of $\beta_{\rm SG}$ purely from the revealed entries. 

In the large $r,\epsilon$ regime, both $\beta_{\rm SG},\beta_{R}$ decay to $0$, so that we can further approximate 
\eq{
1 &= \epsilon \lambda(\beta_{\rm SG})
\underset{r,\epsilon\to \infty}{\sim} \epsilon  r\beta_{\rm SG}^2\E[x^2]\E[y^2]\, , \\
1 &= \epsilon \mu(\beta_{\rm R}) 
\underset{r,\epsilon\to \infty}{\sim} \epsilon \beta_{\rm R}\sqrt{\E[x^2]\E[y^2]} \, ,
}
so that we reach the simple asymptotic expression, in the large $\epsilon,r$ limit, that $\epsilon_c = r$, or equivalently $C(r) = 1$.
It is interesting to note that this result was obtained as the detectability threshold in completion of rank $r=O(n)$ matrices from $O(n^2)$ entries in the Bayes optimal setting in \cite{kabashima2014phase}. Notice, however, that {\it exact} completion in the setting of \cite{kabashima2014phase} is only possible for $\epsilon> r (m+n)/\sqrt{n m}$: clearly detection and exact completion are different phenomena.

\subsection{Computation of the spectral density}
In this section, we show how the spectral density of the Bethe Hessian can be computed analytically on tree-like graphs such as those generated by picking uniformly at random the observed entries of the matrix $XY^{\dag}$.  The spectral density is defined as 
\eq{
\nu(\lambda)=\lim_{n,m\to \infty} \frac{1}{n+m}\overset{n+m}{\underset{i=1}{\sum}}\
\delta(\lambda-\lambda_i)\, , } where the $\lambda_i$'s are the
eigenvalues of the Bethe Hessian. Using again the cavity method, It can be shown
\cite{rogers2008cavitySym} that the spectral density (in
which potential delta peaks have been removed) is given by \eq{
\label{specDens}
\nu(\lambda)=\lim_{n,m\to \infty}\frac{1}{\pi (n+m)}\overset{n+m}{\underset{i=1}{\sum}}\
\text{Im}\Delta_i(\lambda)\, ,
}
where the $\Delta_i$ are complex variables living on the vertices of the graph $\mathcal{G}$, which are given by:
\eq{
\label{full}
\Delta_i=\Big(-\lambda+1 + \underset{ k\in\partial i }{ \sum } \sinh^2{ \beta A_{ik}}-\underset{l\in \partial
i}{\sum}\frac{1}{4}\sinh^2(2\beta A_{il})\Delta_{l\rightarrow i}\Big)^{-1}\, ,
}
where $\partial i$ is
the set of neighbors of $i$. The $\Delta_{i\rightarrow j}$ are the
(linearly stable) solution of the following belief propagation
recursion:
\eq{
\label{BPrec}
\Delta_{i\rightarrow j}=\Big(-\lambda+1 + \underset{ k\in\partial i }{ \sum }  \sinh^2{ \beta A_{ik}}-\underset{l\in \partial
i\backslash j}{\sum}\frac{1}{4}\sinh^2(2\beta A_{il})\Delta_{l\rightarrow i}\Big)^{-1}\, . } The ingredients to derive this formula are to turn the computation of the spectral density into a marginalization problem for a graphical model on the graph $\mathcal{G}$, and then write the belief propagation equations to solve it. Quite remarkably, it has been shown \cite{RSA:RSA20313} that this approach leads to an asymptotically exact (and {\it rigorous}) description of the spectral density on Erd\Horig{o}s-R\'enyi random graphs, which are locally tree-like in the limit where $n,m\rightarrow\infty$. We can again solve equation (\ref{BPrec}) numerically using the belief propagation algorithm. The results are shown on Fig.~\ref{fig:spectrum}: the bulk of the spectrum is always positive.

We now demonstrate that for any value of $\beta<\beta_{\rm SG}$, there exists an open set around $\lambda=0$ where the spectral density vanishes. This justifies independently or choice for the parameter $\beta$. The proof follows \cite{saade2014spectral} and begins by noticing that $\Delta_{i\rightarrow j} = \cosh^{-2}(\beta A_{ij})$ is a fixed point of the recursion (\ref{BPrec}) for $\lambda = 0$. Since this fixed point is real, the corresponding spectral density is $0$. Now consider a small perturbation $\delta_{ij}$ of this solution such that $\Delta_{i\rightarrow j} = \cosh^{-2}(\beta A_{ij})(1+ \cosh^{-2}(\beta A_{ij})\delta_{ij})$. The linearized version of (\ref{BPrec}) writes  $\delta_{i\rightarrow j} = {\sum}_{l\in\partial i\backslash j} \tanh^2(\beta A_{il})\delta_{i\rightarrow l}$ .
The linear operator thus defined is a weighted version of the non-backtracking matrix of \cite{krzakala2013spectral}. Its spectral radius is given by $\rho = \epsilon \lambda(\beta)$, where $\lambda$ is defined in \ref{lambda}. In particular, for $\beta<\beta_{\rm SG}$, $\rho<1$, so that a straightforward application \cite{saade2014spectral} of the implicit function theorem allows to show that there exists a neighborhood $U$ of $0$ such that for any $\lambda\in U$, there exists a real, linearly stable fixed point of (\ref{BPrec}), yielding a spectral density equal to $0$.

\section{Numerical tests}
\label{sec:numerical}

The algorithm was implemented in Julia \cite{BEKS14}, using the NLopt optimization package \cite{johnson2014nlopt} for the minimization of the discrepancy (\ref{costFunction}). A matlab demo using the implementation of the limited-memory BFGS algorithm of \cite{schmidt2005minfunc} is also available. Both demos can be downloaded from \cite{website}.

\begin{figure}[t]
\begin{center}
\includegraphics[width=1\linewidth]{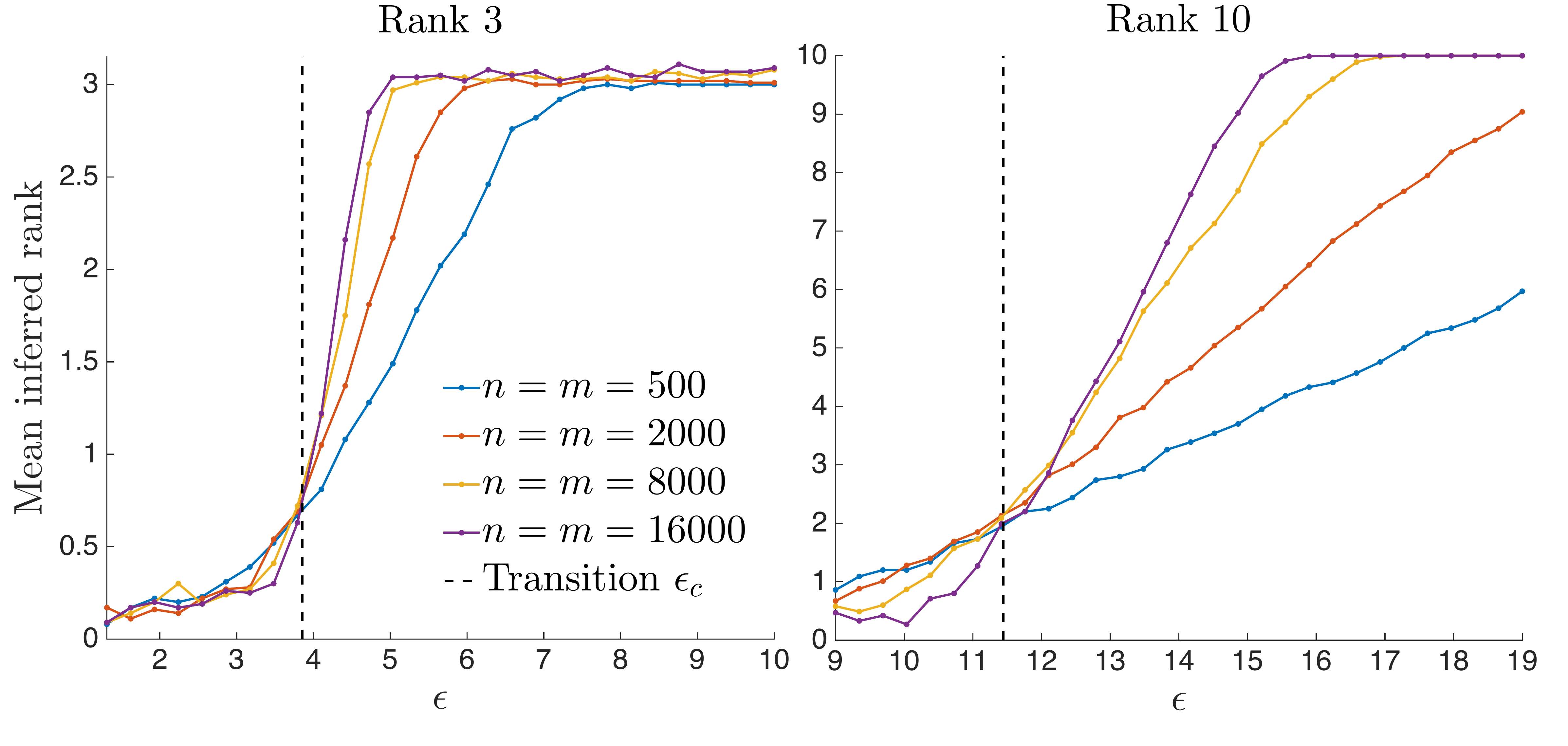}
\end{center}
\caption{Mean inferred rank as a function of $\epsilon$, for different sizes. Each point is averaged over $100$ samples of matrices $XY^{\dag}$ of size $n\times m$, with the entries of $X,Y$ drawn from a Gaussian distribution of mean $0$ and variance~$1$. The theoretical transition is computed with a population dynamics algorithm (see section \ref{analysis}). The finite size effects are considerable but consistent with the asymptotic prediction.    
\label{fig:infer_rank}}
\end{figure}

\begin{figure}[t]
\begin{center}
\includegraphics[width=0.9\linewidth]{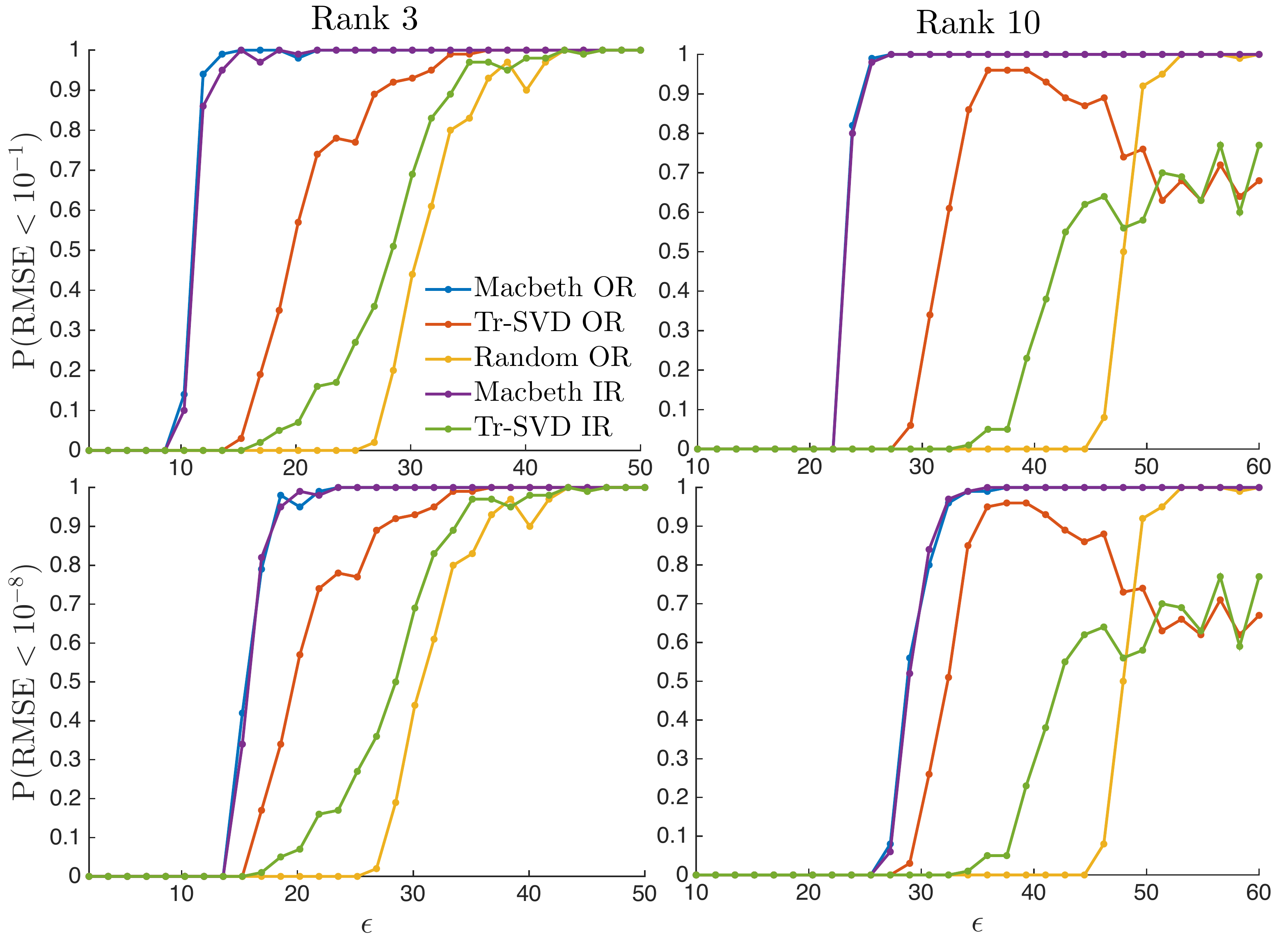}
\end{center}
\caption{RMSE as a function of the number of revealed entries per row $\epsilon$: comparison between different initializations for the optimization of the cost function (\ref{costFunction}). The top row shows the probability that the achieved RMSE is smaller than $10^{-1}$, while the bottom row shows the probability that the final RMSE is smaller than $10^{-8}$. The probabilities were estimated as the frequency of success over $100$ samples of matrices $XY^{\dag}$ of size $10000\times10000$, with the entries of $X,Y$ drawn from a Gaussian distribution of mean $0$ and variance~$1$. All methods optimize the cost function (\ref{costFunction}) using a limited-memory BFGS algorithm \cite{liu1989limited} part of NLopt \cite{johnson2014nlopt}, starting from different initial conditions. The maximum number of iterations was set to $1000$. The initial conditions compared are MaCBetH with oracle rank (MaCBetH OR) or inferred rank (MaCBetH IR), SVD of the observed matrix ${\cal M}$ after trimming, with oracle rank (Tr-SVD OR), or inferred rank (Tr-SVD IR, note that this is equivalent to OptSpace \cite{kmo10} in this regime), and random initial conditions with oracle rank (Random OR). 
For the Tr-SVD IR method, we inferred the rank from the SVD by looking for an index for which the ratio between two consecutive eigenvalues is minimized, as suggested in \cite{keshavan2009low}. 
\label{fig:RMSE}}
\end{figure}

Figure \ref{fig:infer_rank} illustrates the ability of the Bethe Hessian to infer the rank above the critical value~$\epsilon_c$ in the limit of large size $n,m$ (see section \ref{analysis}).  
In Figure \ref{fig:RMSE}, we demonstrate the suitability of the eigenvectors of the Bethe Hessian as starting point for the minimization of the cost function (\ref{costFunction}). We compare the final RMSE achieved on the reconstructed matrix $XY^{\dag}$ with $4$ other initializations of the optimization, including the largest singular vectors of the trimmed matrix ${\cal M}$ \cite{kmo10}. MaCBetH systematically outperforms all the other choices of initial conditions. Remarkably, the performance achieved by MaCBetH with the inferred rank is essentially the same as the one achieved with an oracle rank. By contrast, estimating the correct rank from the (trimmed) SVD is more challenging. We note that for the choice of parameters we consider, trimming had a negligible effect. 
Along the same lines, OptSpace \cite{kmo10} uses a different minimization procedure, but from our tests we could not see any difference in performance due to that. 

\section{Conclusion}

In this paper, we have presented MaCBetH, an algorithm for matrix completion that is efficient for two distinct, complementary, tasks: (i) it has the ability to estimate a finite rank $r$ reliably from fewer random entries than other existing approaches, and (ii) it gives lower root-mean-square reconstruction errors than its competitors. The algorithm is built around the Bethe Hessian matrix and leverages both on recent progresses in the construction of efficient spectral methods for clustering of sparse networks \cite{krzakala2013spectral,saade2014spectral,blm15}, and on the OptSpace approach \cite{kmo10} for matrix completion. Demos in Julia and matlab are available for download \cite{website}.

The method presented here offers a number of possible future directions, including replacing the minimization of the cost function by a message-passing type algorithm, the use of different neural network models, or a more theoretical direction involving the computation of information theoretically optimal transitions for detectability.

\section*{Acknowledgment}
The research leading to these results has received funding from the
European Research Council under the European Union's $7^{th}$
Framework Programme (FP/2007-2013/ERC Grant Agreement 307087-SPARCS).


\bibliographystyle{IEEEtran}
\small{
\bibliography{mybib}
}
\end{document}